\documentclass[prb,aps,twocolumn,superscriptaddress,longbibliography,floatfix,notitlepage]{revtex4-1}
\usepackage{dcolumn}
\usepackage{bm}
\usepackage[english]{babel}
\usepackage[T1]{fontenc}
\usepackage{lmodern}
\usepackage{amsmath,amsfonts,amssymb,amsthm,bm,times,dcolumn}
\usepackage{mwe}
\usepackage{microtype}
\usepackage{braket}
\usepackage{mathrsfs}
\usepackage{nicematrix}
\usepackage{array}
\usepackage{tikz}
\usepackage{blkarray}
\usepackage[colorlinks={true},citecolor={blue},filecolor={blue},linkcolor={blue},urlcolor={blue}]{hyperref}
\usepackage{graphicx,epstopdf,color}
\usepackage{subfigure}
\usepackage[mathscr]{euscript}
\usepackage{setspace}
\usepackage{physics}

\begin{document}

\title{Many-body quantum thermal machines in a Lieb-kagome Hubbard model}

\author{Saikat Sur}
\affiliation{Department of Chemical and Biological Physics \& AMOS,
Weizmann Institute of Science, Rehovot 7610001, Israel}

\author{Pritam Chattopadhyay}
\affiliation{Department of Chemical and Biological Physics \& AMOS,
Weizmann Institute of Science, Rehovot 7610001, Israel}

\author{Madhuparna Karmakar}
\email{madhuparna.k@gmail.com}
\affiliation{Department of Physics and Nanotechnology, College of Engineering and Technology,
SRM Institute of Science and Technology, Kattankulathur, Chennai-603203, India}
\author{Avijit Misra}
\email{avijitmisra0120@gmail.com}
\affiliation{Centre for Quantum Engineering, Research and Education (CQuERE),
TCG CREST, Salt Lake, Sector 5, Kolkata 700091, India}

\date{\today}

\begin{abstract}
Quantum many-body systems serve as a suitable working medium for realizing quantum thermal machines
(QTMs) by offering distinct advantages such as cooperative many-body effects, and performance boost at the quantum
critical points. However, the bulk of the existing literature exploring the criticality of many-body systems in the
context of QTMs involves models sans the electronic interactions, which are non-trivial to deal with and require
sophisticated numerical techniques. Here we adopt the prototypical Hubbard model in two dimensions (2D) in
the framework of the line graph Lieb-kagome lattice for the working medium of a multi-functional QTM. We
resort to a non-perturbative, static path approximated (SPA) Monte Carlo technique to deal with the repulsive
Hubbard model. We observe that in a Stirling cycle, in both the interacting and non-interacting limits, the heat engine function
dominates and its performance gets better when the strain is induced from the kagome to the Lieb limit, while
for the reverse the refrigeration action is preferred. Further, we show that the QTM performs better when the difference between the temperatures of the two baths is lower and the QTM reaches the Carnot limit in this regime. Further, we
extensively study the performance of the QTM in the repulsive Hubbard interacting regime where the magnetic
orders come into the picture. We explore the performance of the QTM along the quantum critical points and
in the large interaction limit.  

\end{abstract}

\maketitle

\section{Introduction}
With recent advancements towards miniaturized devices at the nanoscale, there has been an upsurge of activities in modeling and designing quantum thermal devices including quantum heat engines~\cite{scovil1959three,alicki1979quantum,mukherjee_2021_review,eitan_1996,cangemi2023quantum,gelbwaser2015thermodynamics,PhysRevE.87.012140,opatrny2023nonlinear,Misra_2022,Opatrn__2021,myers2022quantum,bhattacharjee2021quantum,pandit2021non,chattopadhyay2021quantum,saha2023harnessing,sur2023quantum,kurizki2022thermodynamics}, refrigerators~\cite{PhysRevLett.108.070604,PhysRevLett.105.130401,Mukhopadhyay_2018,Das_2019,ray2023kerrtype}, diodes and transistors~\cite{Naseem_2020,PhysRevE.99.042121,PhysRevA.103.052613,Joulain_2016}, transformers~\cite{maity2024quantum}. Though the theoretical explorations have been much ahead, there have been successful experimental demonstrations of quantum heat engines (QHE) in trapped ions~\cite{rossnagel2016single,maslennikov2019quantum,lindenfels2019spin}, nuclear magnetic resonance~\cite{peterson2019experimental}, and nitrogen-vacancy centers in diamonds~\cite{klatzow2019experimental} based setups. One of the main motivations of these explorations is to harness quantumness for better functioning of the QHE which includes exploiting non-thermal and squeezed baths~\cite{scully2002extracting,rossnagel2014nanoscale,gardas2015thermodynamic,niedenzu2018quantum}, presence of quantum coherence in the working medium (WM)~\cite{watanabe2017quantum,camati2019coherence}, quantum  measurements~\cite{elouard2017role,elouard2017extracting,cottet2017observing,santos2023pt}. 

Primarily the study of QHE has dealt with a single particle or a few~\cite{kosloff2017quantum}.  Considering many-body (MB) quantum systems as 
WM of a QHE is relatively new. Quantum MB systems have been employed to harness MB quantum correlation~\cite{jaramillo2016quantum}, quantum criticality~\cite{polettini2015efficiency,PhysRevE.98.052124,campisi2016power}, MB localization~\cite{halpern2019quantum}, super-radiance~\cite{hardal2015superradiant}, minimizing friction~\cite{deng2013boosting,campo2014more,beau2016scaling,binder2018thermodynamics}, and investigate the effect of quantum statistics~\cite{bengtsson2018quantum} in QHE. Quantum criticality has been proven to be a useful resource in MB-QHE~\cite{campisi2016power,chen2019interaction}. 

The bulk of the existing literature exploring the criticality of MB in the context of QHE involves models sans the electronic interactions, which are non-trivial to deal with and 
require sophisticated numerical techniques. At the same time, electronic interactions in conjunction with competing correlations and/or geometric frustration of lattices are known to bring forth rich quantum phases and phase transitions~\cite{dagotto_science2005,balents_natphys2010,haule_rmp2011,kawakami_prb2002,kotlier_prl2004,trembley_prl2006,kawakami_prl2008,georges_prx2021}. The aim of this work is to analyze such an interplay between quantum correlations and geometric frustration in the light of quantum thermodynamics, thereby proposing its prospective application as a quantum thermal machine (QTM).

Our starting Hamiltonian corresponds to the prototypical Hubbard model in two dimensions (2D) in the framework of the line graph Lieb-kagome lattice~\cite{janson_prb2021,paiva_prb2023}. The unprecedented control over the engineering and subsequent tuning of these lattices in ultracold atomic gases~\cite{takahashi_sciadv2015,takahashi_prl2017}, optically induced photonic systems~\cite{molina_prl2015,chen_optlett2016,thomson_optlett2015}, artificial lattices engineered through lithography and atomic manipulations~\cite{liljeroth_natphys2017,swart_natphys2017,zhang_prb2016} and more recently in a metal-organic framework (MOF)~\cite{bredas_mathor2022} have provided the required impetus to the search for their potential applications. In the non-interacting limit, this lattice with a three-site unit cell hosts three electronic dispersion bands as two dispersive and one flat band. The position of the electronic bands and the corresponding underlying Brillouin zone provides the Lieb and the kagome lattices with their distinct properties that have been investigated extensively. The impact of the inclusion of electronic interactions in this system shows up as two important observations viz. $(i)$ beyond a critical interaction, magnetic correlation sets in, demarcating the low-temperature phases to be magnetically disordered and ordered, $(ii)$ the electronic transport, controlled by the underlying band structure quantifies the different interaction regimes of the system as metallic and insulating~\cite{mk_kagome_arxiv}. Being a line-graph lattice the band structure between the Lieb and kagome limits can be smoothly evolved by suitable strain engineering, thereby tuning the thermodynamic, spectroscopic, and transport properties of this system.

In the next few sections, we discuss the impact of this strain engineering on electronic dispersion and explore its implications on the function of a QTM. Our primary inference from this work is that by suitable strain engineering one can achieve four working domains such as quantum heat engine (QHE), quantum refrigerator (QR), quantum accelerator (QA), and quantum heater (QH) by running a Stirling cycle~\cite{thomas_2019,Hamedani_Raja_2021,chattopadhyay2019relativistic,chattopadhyay2021bound,chattopadhyay2020non}. The QHE domain is more feasible and its performance gets better when we engineer the strain from the kagome limit to the Lieb limit in the line graph lattice. Conversely, for QR, it is better, if the strain is engineered in the reverse direction. The performance of both QHE and QR gets better when the temperature difference between the two operating baths, hot and cold namely, is lower and the Carnot limit is achieved in this region. This is valid for both the interacting and the non-interacting cases. 

For the interacting case, though a low-temperature difference between the working bath is preferred, unlike the non-interacting case the favorable working temperatures are not close to the zero temperature. Interestingly, at the critical point where the system exhibits a magnetic transition from a paramagnetic metal to an anti-ferromagnetic metal, the QTM performs well through the entire parameter space of the bath temperature, close to zero being the most favorable region.

The article is organized as follows. In Sec. \ref{model}, we analyze the model of the WM, a 2D line graph lattice (Lieb and kagome) having onsite repulsive Hubbard interaction. Then we discuss the Stirling cycle and thermodynamic quantifiers of heat and work and performance bounds of the same in Sec. \ref{cycle}. In Sec. \ref{results}, we discuss the results for both the non-interacting (Sec. \ref{non-inte}) and interacting (Sec. \ref{inte})
cases. Finally, we conclude in Sec. \ref{conclu}.


\section{Model}
\label{model}
\begin{figure*}[t]
    \centering
    \includegraphics[width=0.80\textwidth]{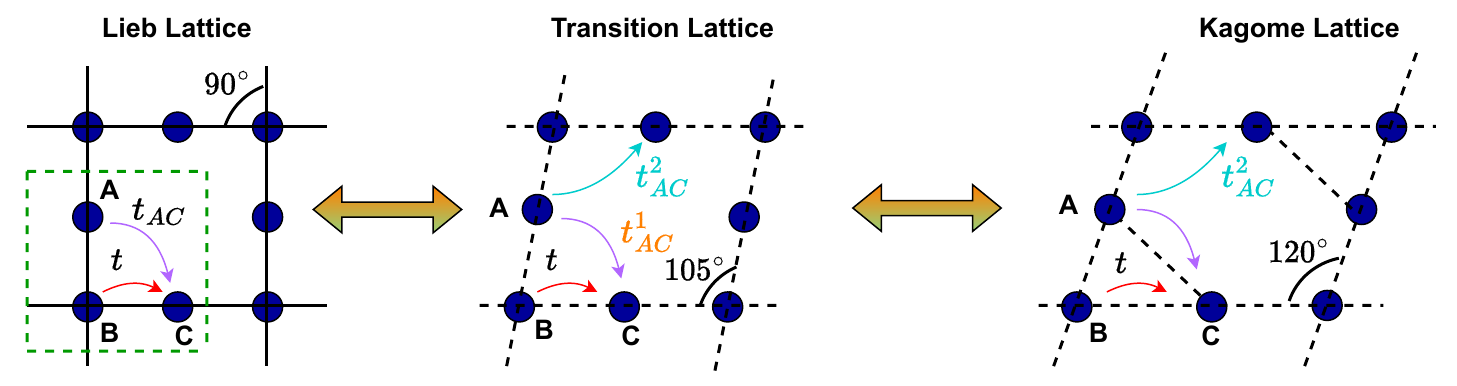}
    \caption{Schematic diagram showing the strain-induced evolution of the lattice structure between the Lieb and the kagome limits.}
    \label{fig1}
\end{figure*}

The prototypical repulsive Hubbard Hamiltonian on a 2D line graph Lieb-kagome lattice reads as
\begin{eqnarray} \label{eq1}    
H & = & \mu \sum_{i, \sigma} c^{\dagger}_{i, \sigma}c_{i, \sigma} + \sum_{\langle ij \rangle, \sigma} t_{ij}  c^{\dagger}_{i, \sigma}c_{j,\sigma} + \sum_{\langle\langle ij \rangle\rangle, \sigma} t'_{ij}  c^{\dagger}_{i, \sigma}c_{j, \sigma} \nonumber \\ && + \mbox{H.c.} + U\sum_{i}\hat n_{i\uparrow}\hat n_{i\downarrow}.
\label{Eq1}
\end{eqnarray}
where, $t_{ij}$ corresponds to the nearest neighbor hopping and $t_{ij}=t=1$ sets the reference energy scale of the model. The applied strain is quantified in terms of $t_{ij}^{\prime} = t^{\prime}$. $U > 0$ corresponds to the repulsive Hubbard interaction. We work at the half-filling and the chemical potential $\mu$ is adjusted to achieve the same. The strain-induced excursion of the lattice and therefore the corresponding electronic band structure between the Lieb and the kagome limits is represented in Fig.~\ref{fig1}.

In the presence of interactions ($U \neq 0$), the model is made numerically tractable via Hubbard-Stratonovich (HS) decomposition of the four-fermion term \cite{hs1,hs2}, introducing a vector ${\bf m}_{i}(\tau)$ and a scalar $\phi_{i}(\tau)$ (bosonic) auxiliary fields at each site, which couples to the spin and the charge channels, respectively. The problem is addressed via the static path approximated (SPA) Monte Carlo technique wherein the model is envisaged as an effective spin-fermion model with the random, fluctuating, "classical" background of the auxiliary fields coupled to the free fermions. The $\phi_{i}$ field is treated at the saddle point level as $\phi_{i} \rightarrow \langle \phi_{i}\rangle = \langle n_{i}\rangle U/2$ (where, $\langle n_{i}\rangle$ is the number density of the fermions), while the complete spatial fluctuations of ${\bf m}_{i}$ are retained. The quantum thermodynamic properties are analyzed based on the proper identification of heat and work in the quantum domain \cite{Misra_2015,Skrzypczyk_2014}. Thereby, we analyze the performance of a multifunctional (see Table. \ref{tab:1}) QTM.  The results presented in this paper correspond to a system size of $3\times L^{2}$, with $L=16$, and are verified to be robust against finite system size effects. The details of the numerical technique are presented in the appendix.

We begin the analysis of our results with the non-interacting limit by setting $U=0$ in Eq. \eqref{Eq1}.
To evaluate the energy spectrum, we transform the Hamiltonian in the momentum space as $\mathscr{H} = \sum_k \Phi_k^\dagger H_k \Phi_k$ with $\Phi_k^\dagger = \begin{pmatrix}
c^\dagger_{\textbf{Ak}} & c^\dagger_{\textbf{Bk}} & c^\dagger_{\textbf{Ck}} 
\end{pmatrix}$, where 
\begin{eqnarray}
 H_{\textbf{k}} =   \begin{pmatrix}
0 & A_{\textbf{k}}  &   B_{\textbf{k}}\\
{A_{\textbf{k}}} &  0  & C_{\textbf{k}}\\
{B_{\textbf{k}}} &  C_{\textbf{k}} &  0
\end{pmatrix}.\nonumber\\ 
\label{H_k}
\end{eqnarray}
Here $A_{\textbf{k}} = -2 \cos \frac{k_x}{2}$, $B_{\textbf{k}} = -2 \cos(-\frac{k_x}{2} \cos\theta + \frac{k_y}{2}\sin\theta)$, and $C_{\textbf{k}}= -t^1_{AC} \cos(\frac{k_x}{2}(1+\cos \theta) -\frac{k_y}{2}\sin \theta )- t^2_{AC} \cos(\frac{k_x}{2}(1-\cos \theta) +\frac{k_y}{2}\sin \theta )$.
The strain is quantified in terms of the parameters $t^1_{AC}$ and $t^2_{AC}$ between the sites $A$ and $C$ as~\cite{jiang_2019_prb}
\begin{eqnarray}\label{eqttttt}
   &&t^1_{AC} = \exp[\eta(1-2\cos \frac{\theta}{2})],\nonumber\\
   &&t^2_{AC} = \exp[\eta(1-2\sin \frac{\theta}{2})],
\end{eqnarray}
where the parameter $\eta$ governs the rate at which the 
strain diminishes with distance. A higher $\eta$ results in a more gradual decrease in the 
band curvature, leading to a flatter band. Conversely, a smaller $\eta$ is more adept at capturing the transition of the band structure from Lieb to kagome. For $\eta=8$, we achieve relatively flat bands and smooth transitions between the two lattices~\cite{jiang2019topological}. The angle $\theta$ is tuned within the range of $[\frac{\pi}{2},\frac{2\pi}{3}]$. The lower and the upper bounds correspond to the Lieb and the kagome lattice,  respectively. In the Lieb limit ($\theta = \pi/2$), with the strain being,  $t^1_{AC} = t^2_{AC} = \exp[\eta(1-\sqrt{2})]$. On the other hand, in the kagome limit ($\theta = 2\pi/3$), with the corresponding strain being parametrized as, $t^1_{AC} =t=1$ and $t^2_{AC} = \exp[\eta(1-\sqrt{3})]$. The transition between the Lieb and kagome lattices is observed at $\theta^* = 7\pi/12$.

 Now diagonalizing $H_k$, we get the three respective eigenvalues
\begin{eqnarray}\nonumber
&&\varepsilon_{\mathbf{k},j} = \sqrt{\frac{2}{3}} (A^2_{\textbf{k}}+{B_{\textbf{k}}}^2+{C_{\textbf{k}}}^2)^{1/2} \times \nonumber\\
&& \cos\left[ \frac{1}{3} \cos^{-1} \big(\frac{3\sqrt{3} A_{\textbf{k}}B_{\textbf{k}}C_{\textbf{k}}}{ (A^2_{\textbf{k}}+{B_{\textbf{k}}}^2+{C_{\textbf{k}}}^2)^{3/2}}\big) - \frac{2\pi (j-1)}{3} \right], \nonumber\\
&&{~~} j = 1,2,3.
\label{analytic}
\end{eqnarray}

\section{Stirling cycle and thermodynamic quantities}
\label{cycle}
\begin{figure}[t]
    \centering
     \includegraphics[width=0.45\textwidth]{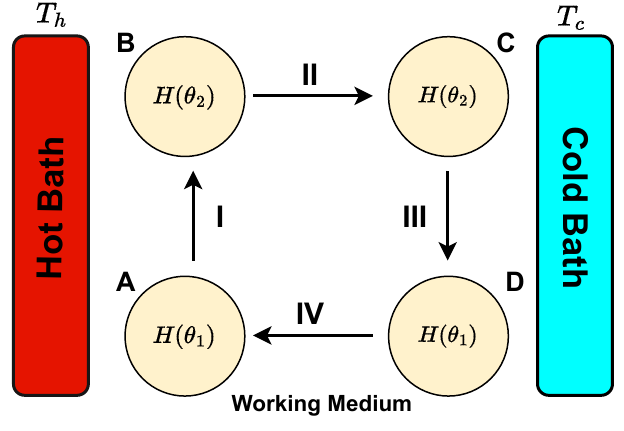}
   
    \caption{Schematic representation of the Stirling cycle. $A\rightarrow B$ and $C\rightarrow D$ are the two isothermal steps where the Hamiltonian parameter $\theta$ is changed when the WM is in contact with the heat baths at temperatures $T_h$ and $T_c$ respectively. $B\rightarrow C$ and $D\rightarrow A$ are the two isochoric thermalization steps in contact with the heat baths at temperatures $T_h$ and $T_c$ respectively, while the Hamiltonian parameter $\theta$ is fixed.}
    \label{Stirling}
\end{figure}
A quantum Stirling cycle~\cite{thomas_2019,Hamedani_Raja_2021,chattopadhyay2019relativistic,chattopadhyay2021bound,chattopadhyay2020non} consists of four steps (Fig. \ref{Stirling}). First $(A \rightarrow B)$, the WM is in equilibrium with the heat bath at temperature $T_h$, while the Hamiltonian parameter is isothermally changed, $\theta_1 \rightarrow \theta_2$ (Eq.~\eqref{eqttttt}).  In the second step $(B \rightarrow C)$, the WM is detached from the hot bath and connected with a cold bath at a temperature $T_c$ with which it thermalizes isochorically without performing any mechanical work. Next $(C \rightarrow D)$, the WM is brought back to its initial Hamiltonian configuration isothermally ($\theta_2 \rightarrow \theta_1$) while keeping the system connected to the cold bath. In the final step $(D \rightarrow A)$, the system is detached from the cold bath and connected to the hot bath where it undergoes an isochoric thermalization process without performing any mechanical work. Thus, the WM fully recovers to its initial state and the cycle is completed.

The total heat transferred to the system from the bath in the isothermal processes $(A \rightarrow B)$ and $(C \rightarrow D)$, while keeping the system in equilibrium with the bath at temperature $T_h$ and $T_c$ respectively, are combinations of the mechanical works performed due to deformation of the potential and the change of the internal energies. Thus the heat transferred to the bath  from the system during two isothermal processes is given by
\begin{eqnarray}
Q_{AB} = -(U^{\theta_2}_{T_h}-U^{\theta_1}_{T_h}+k_B T_h \ln Z^{\theta_2}_{T_h} - k_B T_h \ln Z^{\theta_1}_{T_h}), 
\end{eqnarray}
and
\begin{eqnarray}
Q_{CD} = -(U^{\theta_1}_{T_c}-U^{\theta_2}_{T_c}+k_B T_c \ln Z^{\theta_1}_{T_c} - k_B T_c \ln Z^{\theta_2}_{T_c}),
\end{eqnarray}
respectively. Here, $Z^\theta_{T}$ is the partition function evaluated as
\begin{eqnarray}
 Z^\theta_{T} = \sum_{\mathbf{k}} \sum_{j}e^{-\frac{\varepsilon^\theta_{\mathbf{k},j}}{T}} = \frac{1}{4\pi^2} \sum_j\int \int dk_x dk_y e^{-\frac{\varepsilon^\theta_{\mathbf{k},j}}{T}}
\end{eqnarray}
and the internal energy $U^\theta_{T}$ is evaluated as $T^2 \frac{\partial}{\partial T} \ln  Z^\theta_{T}.$
 On the other hand, the heat transferred to the system during the isochoric thermalization processes $(B \rightarrow C)$ and $(D \rightarrow A)$ are the differences between the average energies of the initial and the final configurations as no mechanical work has been considered. Therefore, we have
\begin{eqnarray}
Q_{BC} = -(U^{\theta_2}_{T_c}-U^{\theta_2}_{T_h}),\;\; 
\text{and}\;\; Q_{DA} = -(U^{\theta_1}_{T_h}-U^{\theta_1}_{T_c}).
\end{eqnarray}

Provided all the processes involved in the cycle are reversible and no leakage takes place, the expressions for the net work done on the system $W$ and the heat transferred to the system with the hot and cold baths, $Q_h$ and $Q_c$ after completion of one cycle are given as follow:
\begin{eqnarray}
  W &=& (Q_{AB} + Q_{BC} +Q_{CD} + Q_{DA}) \nonumber\\&&= - T_h\ln \frac{Z^{\theta_2}_{T_h}}{Z^{\theta_1}_{T_h}} + T_c \ln \frac{Z^{\theta_2}_{T_c}}{Z^{\theta_1}_{T_c}},\nonumber\\
  Q_h &=&  -(Q_{AB} + Q_{DA}) = T_h\ln \frac{Z^{\theta_2}_{T_h}}{Z^{\theta_1}_{T_h}}+ U^{\theta_2}_{T_h} -U^{\theta_1}_{T_c}, \nonumber\\
  Q_c &=& -(Q_{BC} + Q_{CD}) = -T_c \ln \frac{Z^{\theta_2}_{T_c}}{Z^{\theta_1}_{T_c}}-U^{\theta_2}_{T_h} +U^{\theta_1}_{T_c}. \nonumber\\
  \label{eq:w}
\end{eqnarray}

By the principle of conservation of energy or the first law of thermodynamics, one can check that $Q_h +Q_c+W = 0$. Importantly, it has been established that the maximal isothermal work extraction also in the quantum domain is given by the change in free energy $\Delta F$~\cite{Misra_2015,Skrzypczyk_2014}, where 
\begin{eqnarray}
    F=U-TS.
\end{eqnarray}
Here, $S$ is the von Neuman entropy of the WM $\rho$ as quantified by $S(\rho)=-\mbox{Tr}[\rho \log \rho]$ and $T$ is the temperature of the heat bath. One can check that work $W$ in Eq. (\ref{eq:w}) follows it.
If the quantities $Q_h$ or $Q_c$ are positive,  the heat is flowing into the system, similarly, if work $W$ is positive, work is done on the system. In conformity with the second law of thermodynamics, only four modes of operation are possible. The four possible modes can be identified by the signs of $W$, $Q_h$, and $Q_c$, as depicted in Table~\ref{tab:1}.  

\vskip 1em
\begin{table}[ht]
\centering
\begin{tabular}{ |p{4cm}||p{1cm}|p{1cm}|p{1cm}|  }
 \hline
Modes of operation & $Q_h$ & $Q_c$ & $W$ \\
 \hline
{ Engine}  & $>0$ & $<0$ &  $<0$\\
 { Refrigerator}&   $<0$  & $>0$   & $>0$\\
{ Accelerator }   & $>0$    & $<0$&   $>0$\\
 { Heater}  & $<0$ & $<0$&  $>0$\\
  \hline
\end{tabular}
\caption{Different modes of operation of a quantum thermodynamic cycle.}
\label{tab:1}
\end{table}

The efficiency $\eta$ of the engine is given by 
\begin{equation}
    \eta= \frac{|W|}{Q_h},
\end{equation}
and the coefficient of the performance (COP) of the refrigerator is given as 
\begin{eqnarray}
 \eta_R = \frac{Q_c}{|W|}.
  \label{eq:eta_R}
\end{eqnarray}
Given a pair of thermal baths, the maximum efficiency or coefficient of performance is obtained for the Carnot cycle, a theoretical proposition with only reversible cycles. We present the efficiency of the engine and coefficient of performance of the refrigerator scaled by the same as its Carnot counterpart in this article.   The efficiency of a Carnot engine and the coefficient of performance of the Carnot refrigerator are given respectively as

\begin{eqnarray}
 \eta^{max} = 1 - \frac{T_c}{T_h}, \mbox{~~and ~~}
\eta^{max}_R = \frac{T_c}{T_h - T_c}.
\end{eqnarray}

\section{Performance of the many-body quantum thermal machine}
\label{results}

\begin{figure}
 \includegraphics[width=0.50\textwidth]{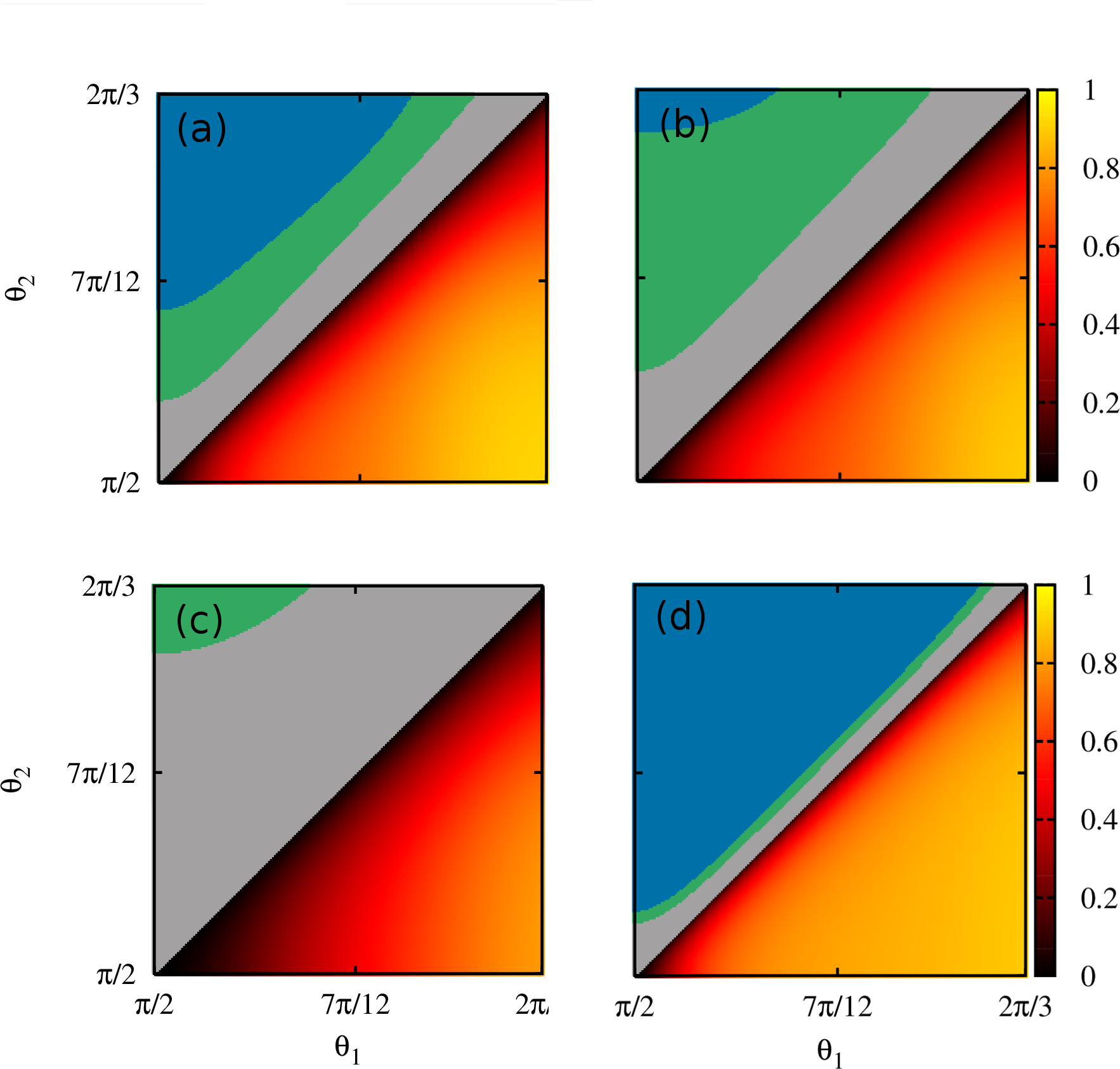}
 \caption{Demonstration of multi-functionality of the  Stirling cycle by tuning the WM strain $(\theta_1,\theta_2)$ without the onsite repulsive interaction, with different sets of bath temperatures: (a) $T_h = 0.02, T_c = 0.01$; (b) $T_h = 0.03, T_c = 0.01$; (c) $T_h = 0.2, T_c = 0.01$; (d) $T_h = 0.05, T_c = 0.04$. The refrigerator, the accelerator, and the heater regions are marked in blue, grey, and green respectively. The engine region is plotted with color density where the color denotes the engine efficiency.}
  \label{fig-theta} 
\end{figure}

In this section, we investigate and analyze the thermodynamic performance of the many-body quantum thermal machine based on the Lieb-kagome Hubbard model. Interestingly, we observe that all four operational regimes are mentioned in the Table. \ref{tab:1}, i.e. QHE, QR, QA, and QH can be realized in our model by either tuning the strain, the bath temperatures, or the onsite repulsive Hubbard interaction. Thus, the model works as a multifunctional quantum thermal device. For brevity, we present the analysis for the non-interacting ($U=0$) and interacting ($U\neq 0$) WM in separate subsections.

\subsection{For non-interacting working medium}
\label{non-inte}

\begin{figure}
 \includegraphics[width=0.51\textwidth]{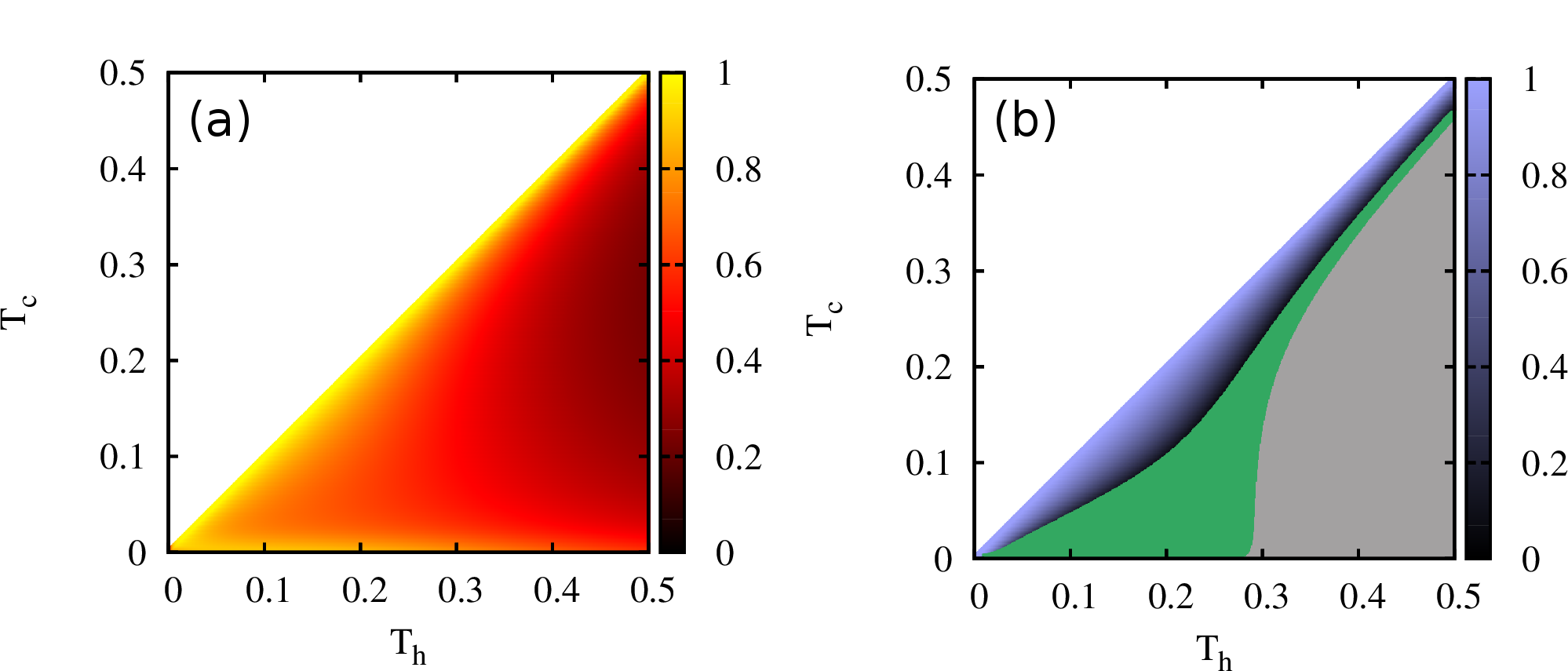}
 \caption{Function of the Stirling cycle with two fixed strain configurations (a)  $\theta_1 = 2\pi/3, \theta_2 = \pi/2$; (b)  $\theta_1 = \pi/2, \theta_2 = 2\pi/3$  as a function of the thermal bath temperatures. The refrigerator, the accelerator, and the heater regions are marked in blue, grey, and green respectively. The engine and refrigerator region is plotted with color density where the color denotes the efficiency and coefficient of performance (COP) of the engine and refrigerator respectively.}
  \label{fig-theta-temp} 
\end{figure}

We begin with the Stirling cycle by tuning the strain of the WM $(\theta_1,\theta_2)$. The performance of the QHE and other functional regions QR, QA, and QH have been shown in Fig. \ref{fig-theta}. It shows that we get the QHE performance better when we start the cycle from the Kagome limit and then move to the Lieb limit and the opposite for QR. The maximum efficiency of the QHE  reaches the Carnot bound for the Lieb-kagome configuration, specifically when $\theta_1 = 2\pi/3$ and $\theta_2 = \pi/2$. In the low-temperature limit, the performance of the QHE and feasible region (in the strain parameter space) for it and QR increases when the difference between the two bath temperatures is lower. Moreover, we observe distinct signatures of all the functions of the proposed QTM. However, as the temperature difference increases, the characteristic features of the QR and QH diminish. 

We further look into the role of temperature difference in Fig. \ref{fig-theta-temp} to substantiate our observation. It clearly shows that the efficiency or the coefficient of performance (COP) of the QHE or QR are higher near the $T_h=T_c$ line and for the configuration with $T_c \rightarrow 0$, $T_h\rightarrow \epsilon$ (where $\epsilon$ is close to zero) with fixed strain configuration. 
 
 
\subsection{For interacting working medium}
\label{inte}

\begin{figure}
\includegraphics[height=5cm,width=8cm,angle=0]{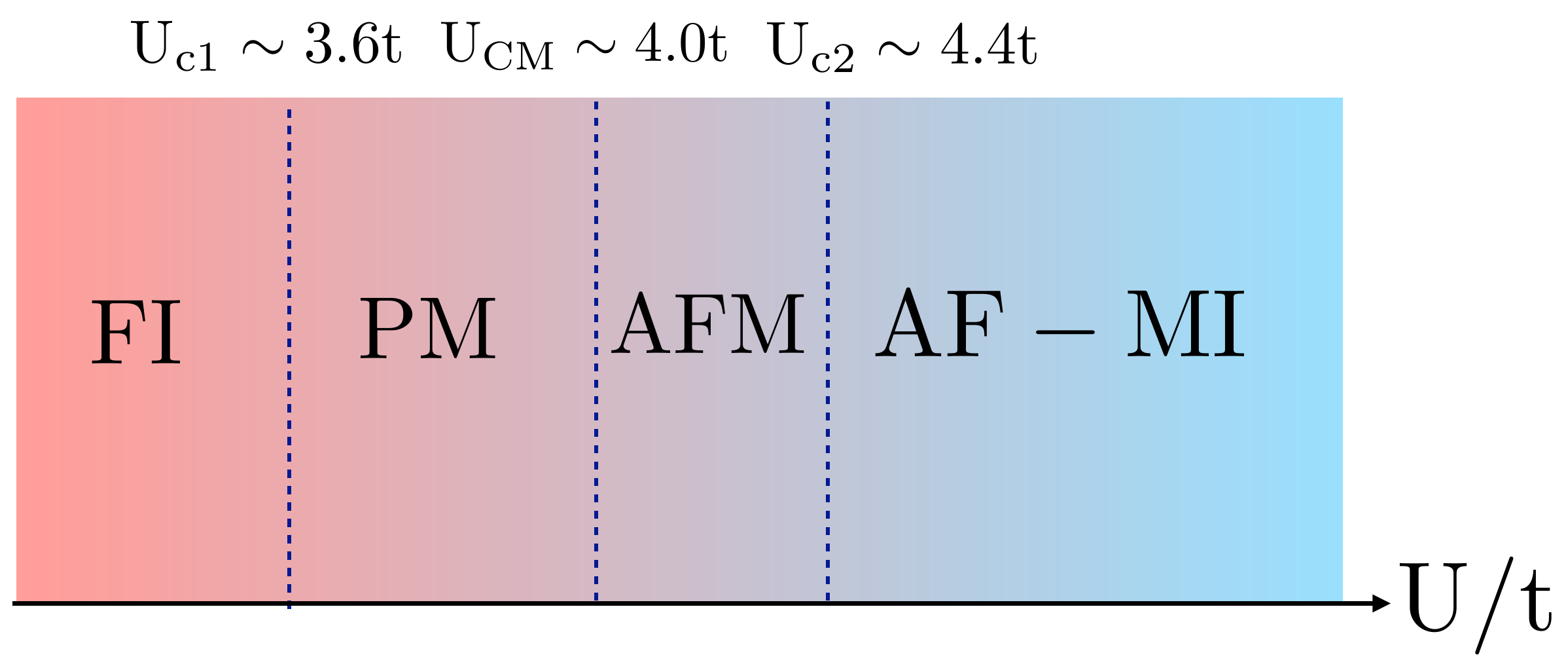}
\caption{Low-temperature phase diagram (at $T = 0.01$) of the kagome Hubbard model at half filling mapping out the various thermodynamic phases as (i) Flat band Insulator (FI), (ii) Paramagnetic Metal (PM), (iii) Antiferromagnetic Metal (AFM) and (iv) Antiferromagnetic Mott Insulator (AF-MI)\cite{mk_kagome_arxiv}.}
\label{fig-phases}
\end{figure}

Let us now explore the interacting WM. The interacting domain hosts several critical points.
In the kagome limit the ground state phase diagram hosts several quantum critical points, as shown in Fig.~\ref{fig-phases}.
Here, the QPT $U_{CM}$ corresponds to a magnetic transition between a 
paramagnetic and an antiferromagnetic metal. $U_{c1}$ shows an insulator-metal 
transition (IMT), while $U_{c2}$ corresponds to a metal-Mott insulator transition (MIT). The phases and transitions are determined based on various thermodynamics,  spectroscopic, and electrical/optical transport properties, viz. magnetic correlations, structure factor, single-particle density of states, optical and electrical conductivity, etc. 

These phases and phase transitions can also be captured based on the quantum information perspective based on indicators such as entanglement entropy, mutual information, etc. and such studies have been reported for a square lattice that has a relatively simple phase 
diagram~\cite{trembley_pnas2021,trembley_prxq2020,trembley_prb2019,trembley_prl2019,yang_prb2023}. The several standard Green's function-based techniques are used to calculate the relevant quantities viz. determinant quantum Monte Carlo, dynamical mean field theory, static path approximated quantum Monte Carlo, etc.

\begin{figure}
\includegraphics[width=0.48\textwidth]{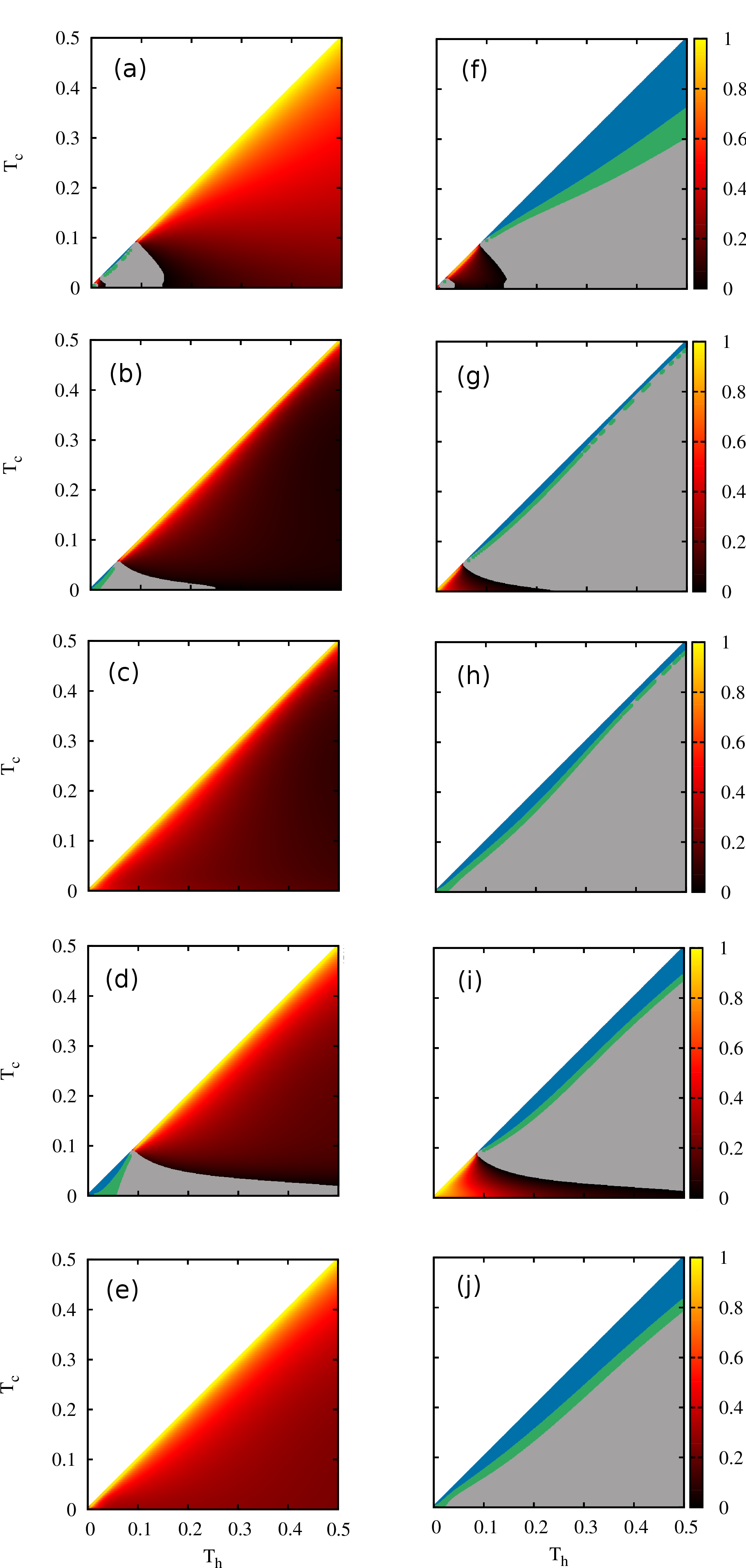} 
 \caption{The diagram of modes for a Stirling cycle operating in the refrigerator, the accelerator, and the heater regions are marked in blue, grey, and green respectively. The set of subfigures (a-e) corresponds to the Stirling cycle with strain parameters $(\theta_1 = 2\pi/3, \theta_2 = \pi/2)$ and the set (f-j) correspond to the same with strain parameters $(\theta_1 = \pi/2, \theta_2 = 2\pi/3)$. The Hubbard interaction parameter assumes the following values:  $U = 1.0$ in (a) and (f); $U = 3.6$ in (b) and (g); $U = 4.2$ in (c) and (h); $U = 4.6$ in (d) and (i); 
 $U = 9.0$ in (e) and (j). The engine region is plotted with color density where the color denotes the engine efficiency.}
 \label{fig_2} 
 \end{figure}

We demonstrate the behavior of the Stirling cycle with different values of the Hubbard interaction parameter computed from numerical analysis in Fig~\ref{fig_2} (a-j) (see App.~B for details regarding the analysis). We have chosen some representative values of $U$ to display the changing of QHE, QR, QA, and QH and their relationship with the quantum phases in the model. Although introducing the interaction in the model does not drastically change the behavior of the cycle for a given set of strain parameters (cf Fig.~\ref{fig-theta-temp}), it is observed that the regions corresponding to different modes do change non-trivially showing the signature of various phases of the system. 

For a cycle with the strain parameters  $(\theta_1 = 2\pi/3, \theta_2 = \pi/2)$, the regions of QH, QR and QA (where $W>0$) exist in the mode-diagrams for $U= 1.0, 3.6$ and $U=4.6$ (see Fig~\ref{fig_2} (a), (b) and (d)) but missing for $U = 4.6$ and $9.0$ when the bath temperatures are low. On the other hand, for a cycle with the strain parameters  $(\theta_1 = \pi/2, \theta_2 = 2\pi/3)$, a small region of QHE appears in the mode diagram for $U= 1.0, 3.6$ and $U=4.6$ when the bath temperatures are low (see Fig~\ref{fig_2} (f), (g) and (i)). Otherwise, in these cases, most of the mode-diagram is still dominated by the QHE mode (where $W<0$) and QH, QA, QR modes (where $W>0$) for the set of strain parameters  $(\theta_1 = 3\pi/2, \theta_2 = \pi/2)$ and   $(\theta_1 = \pi/2, \theta_2 = 2\pi/3)$ respectively, as already shown for the noninteracting counterpart.

\begin{figure}
 \includegraphics[width=0.50\textwidth]{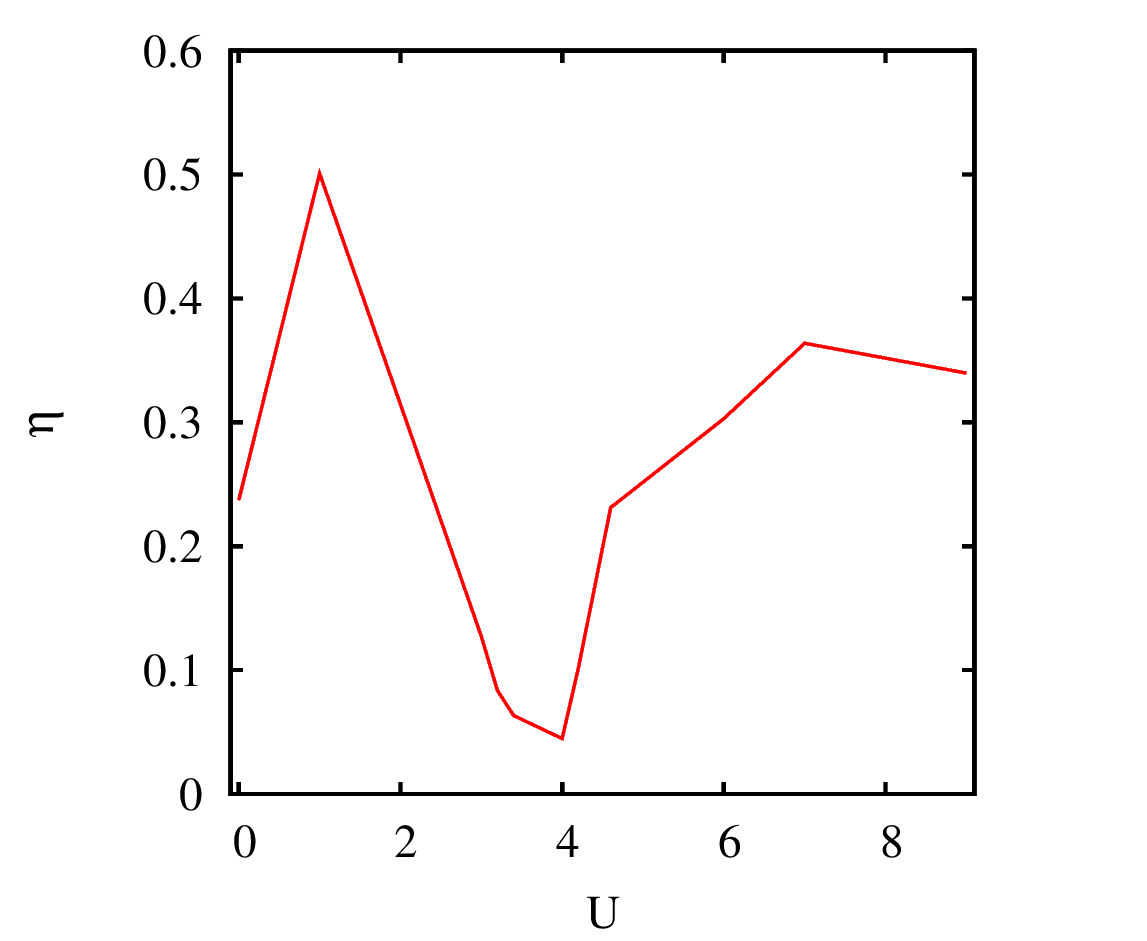}
 \caption{ Efficiency ($\theta_1 = 2\pi/3, \theta_2 = \pi/2$) as a function of $U$ for the set of bath temperatures $T_h = 0.5, T_c = 0.3$.}
 \label{fig_4} 
 \end{figure}

This non-triviality of the results with the interacting model emerges as an artifact of the non-monotonic change in the energy levels of the system in both strain configurations. In other words, the behavior of the cycle will be determined by the phase diagrams of the model with two different strain configurations. In general, we observe that introducing the interaction in the model non-trivially changes the efficiency of the engine for a given pair of bath temperatures as shown in Fig.~\ref{fig_4}. The efficiency first increases with the interaction, then decreases and increases further. It is observed to be minimal near the regions of phase transitions.

\section{Conclusions} 
\label{conclu}

Quantum many-body (MB) systems have been proven to be useful resources for realizing quantum thermal machines (QTMs) as they offer several benefits such as cooperative performance boost, simultaneous achievements of high efficiency and power at the quantum critical points, minimizing frictional effects, and many more. However, the bulk of the existing literature exploring the criticality of MB in the context of QTMs involves models sans the electronic interactions, which are non-trivial to deal with and require
sophisticated numerical techniques. At the same time, electronic interactions in conjunction with competing correlations
and/or geometric frustration of lattices are known to bring forth
rich quantum phases and phase transitions. To harness quantum criticality for QTMs, here we resort to the prototypical Hubbard model in two dimensions (2D) in
the framework of the line graph Lieb-kagome lattice as the working medium of a multi-functional QTM. We
resort to a non-perturbative, static path approximated (SPA) Monte Carlo technique to deal with the repulsive
Hubbard model. We demonstrate that with this model operating between the hot and cold baths one can achieve multiple thermal actions like quantum heat engines, refrigerators, accelerators, and heaters in a quantum Stirling cycle by either engineering the system strain or tuning the bath temperatures. 

Furthermore, we show that the heat engine action gets better when the strain is engineered from the kagome limit to the Lieb limit and the refrigeration gets better for the converse. For both cases, in the low-temperature limit, we show that the proposed QTM reaches the ultimate Carnot limit when the temperature difference between the baths is low. The introduction of the onsite Hubbard repulsion allows one to have the Carnot efficiency even in the relatively high-temperature limit while the difference between the two bath temperatures is still close to zero.  It indicates that a relatively higher temperature than the absolute zero is preferred for intermediate interaction strength $U$ while operating a heat engine by engineering the strain from the kagome to the Lieb limit and the converse while operating a refrigerator, near the interaction strength ($U=4.0t$) for which we get a paramagnetic metal to an anti-ferromagnetic metal, the low-temperature regimes is preferred for both the thermal actions just like the non-interacting case ($U=0$). 

In conclusion, we have modeled a multifunctional QTM, based on a 2D line graph lattice having onsite Hubbard repulsion in the framework of the Lieb and the kagome lattices, which is very efficient. The unprecedented
control over the engineering and subsequent tuning of these
lattices in ultracold atomic gases, optically induced photonic systems, artificial lattices engineered through lithography and atomic manipulations, and more recently in a metal-organic framework (MOF) make the proposed QTM experimentally realizable in the near future. Therefore, our study can be a stepping stone towards exploring the role of strong electronic interaction, which offers rich phases and phase transitions involving magnetic orders, on QTMs.


\onecolumngrid
\appendix
\newpage

\section{Working Medium: Tight-binding calculation \label{Working}}

The electron field operator in the momentum space can be expressed in the quasiparticle basis via Bogoliubov 
canonical transformation as,
\begin{eqnarray}
c_{i\sigma,k} = \sum_{n}u_{i\sigma}^{n}\gamma_{n,k} \\ 
c_{i\sigma,k}^{\dagger} = \sum_{n}u_{i\sigma}^{n*}\gamma_{n,k}^{\dagger}
\end{eqnarray}
where, $\sigma = \pm 1$ corresponds to the spin label and $n$ is the energy index. 
The operators create (annihilate) a Bogoliubov quasiparticle at state $n$ in the momentum space. The quasiparticle operators satisfy the anti-commutation relations:
\begin{eqnarray}
\{\gamma_{n,k}, \gamma_{m,k}^{\dagger}\} = \delta_{nm} \\ 
\{\gamma_{n,k}, \gamma_{m,k}\} = \{\gamma_{n,k}^{\dagger}, \gamma_{m,k}^{\dagger}\} = 0
\end{eqnarray}
These relations also guarantee the anti-commutation relations among the original electronic field operators. With the above canonical transformation, the Hamiltonian is diagonalized in the following form:
\begin{eqnarray}
H_{k}^{\text{diag}} &=& \sum_{n}E_{n}\gamma_{n,k}^{\dagger}\gamma_{n,k} + E_{const}     
\end{eqnarray}
The above Hamiltonian is diagonal in the quasiparticle basis in the momentum space. 
So, the eigenvalues of the Hamiltonian for a $k$ mode given in Eq.~\ref{H_k} are given by Vi\'ete's formula~\cite{nickalls2006},
\begin{eqnarray}\nonumber
&&\varepsilon^\theta_{\mathbf{k},j} = \sqrt{\frac{2}{3}} (A^2_{\textbf{k}}+{B^\theta_{\textbf{k}}}^2+{C^\theta_{\textbf{k}}}^2)^{1/2} \times \nonumber\\
&& \cos\left[ \frac{1}{3} \cos^{-1} \big(\frac{3\sqrt{3} A_{\textbf{k}}B^\theta_{\textbf{k}}C^\theta_{\textbf{k}}}{ (A^2_{\textbf{k}}+{B^\theta_{\textbf{k}}}^2+{C^\theta_{\textbf{k}}}^2)^{3/2}}\big) - \frac{2\pi (j-1)}{3} \right], \nonumber\\
&&{~~} j = 1,2,3.
\end{eqnarray}




\section{Numerical scheme for interacting Hamiltonian \label{Numerical}}

The repulsive Hubbard Hamiltonian on a 2D line graph Lieb-kagome lattice reads as, 
\begin{eqnarray} \label{eq1}    
H & = & \mu \sum_{i, \sigma} c^{\dagger}_{i, \sigma}c_{i, \sigma} + \sum_{\langle ij \rangle, \sigma} t_{ij} c^{\dagger}_{i, \sigma}c_{j,\sigma} + \mbox{H. c.} + \sum_{\langle\langle ij \rangle\rangle, \sigma} t'_{ij}  c^{\dagger}_{i, \sigma}c_{j, \sigma}+ \mbox{H. c.} \nonumber \\ && + U\sum_{i}\hat n_{i\uparrow}\hat n_{i\downarrow}.
\label{Eqn1}
\end{eqnarray}
where, $t_{ij}$ corresponds to the nearest neighbor hopping and $t_{ij}=t=1$ sets the reference energy scale of the model. The applied strain is quantified in terms of $t_{ij}^{\prime} = t^{\prime}$. $U > 0$ corresponds to the repulsive Hubbard interaction. We work at the half-filling and the chemical potential $\mu$ is adjusted to achieve the same.
In order to make the model numerically tractable we decompose the interaction term using Hubbard Stratonovich 
(HS) decomposition and thereby introduce two (bosonic) auxiliary fields viz. a vector field ${\bf m}_{i}(\tau)$ and a scalar field $\phi_{i}(\tau)$, which couples to the spin and charge densities, respectively. The introduction of these auxiliary fields aids in capturing the Hartree-Fock theory at the saddle point and retains the spin rotation invariance and the Goldstone modes. 

In terms of the Grassmann fields $\psi_{i\sigma}(\tau)$, we have,

\begin{eqnarray}
\exp[U\sum_{i}\bar\psi_{i\uparrow}\psi_{i\uparrow}\bar\psi_{i\downarrow}\psi_{i\downarrow}] & = & \int {\bf \Pi}_{i}\frac{d\phi_{i}d{\bf m}_{i}}{4\pi^{2}U}{\exp}[\frac{\phi_{i}^{2}}{U}+i\phi_{i}\rho_{i} +\frac{m_{i}^{2}}{U} -2{\bf m}_{i}.{\bf s}_{i}]
\end{eqnarray}
where, the charge and spin densities are defined as, $\rho_{i} = \sum_{\sigma}\bar\psi_{i\sigma}\psi_{i\sigma}$ and ${\bf s}_{i}=(1/2)\sum_{a,b}\bar \psi_{ia}{\bf \sigma}_{ab}\psi_{ib}$, respectively. The corresponding partition function thus takes the form,
\begin{eqnarray}
{\cal Z} & = & \int {\bf \Pi}_{i}\frac{d\bar\psi_{i\sigma}d\psi_{i\sigma}d\phi_{i}d{\bf m}_{i}}{4\pi^{2}U}
\exp[-\int_{0}^{\beta}{\cal L}(\tau)]
\end{eqnarray}
where, the Lagrangian ${\cal L}$ is defined as,
\begin{eqnarray}
{\cal L}(\tau) & = & \sum_{i\sigma}\bar\psi_{i\sigma}(\tau)\partial_{\tau}\psi_{i\sigma}(\tau) + H_{0}(\tau) +\sum_{i}[\frac{\phi_{i}(\tau)^{2}}{U}+(i\phi_{i}(\tau)-\mu)\rho_{i}(\tau)+
\frac{m_{i}(\tau)^{2}}{U} -2{\bf m}_{i}(\tau).{\bf s}_{i}(\tau)]
\end{eqnarray}

where, $H_{0}$ is the kinetic energy contribution. The $\psi$ integral is now quadratic but at the cost of an additional integration over
the fields ${\bf m}_{i}(\tau)$ and $\phi_{i}(\tau)$. The weight factor for the ${\bf m}_{i}$ and $\phi_{i}$ configurations can be determined by integrating out the $\psi$ and $\bar \psi$, and using these weighted configurations one goes back and computes the fermionic properties. 
Formally,

{\begin{eqnarray}
{\cal Z} & = & \int {\cal D}{\bf m}{\cal D}{\phi}e^{-S_{eff}\{{\bf m},\phi\}}
\end{eqnarray}}
\begin{eqnarray}
S_{eff} & = & \log Det[{\cal G}^{-1}\{{\bf m},\phi\}] + \frac{\phi_{i}^{2}}{U} +
\frac{m_{i}^{2}}{U}
\end{eqnarray}
where, ${\cal G}$ is the electron Green's function in a $\{{\bf m}_{i},\phi_{i}\}$ background.

The weight factor for an arbitrary space-time configuration $\{{\bf m}_{i}(\tau), \phi_{i}(\tau)\}$ involves computation of the fermionic determinant in that background. The auxiliary field quantum Monte Carlo 
with static path approximation (SPA) retains the full spatial dependence in ${\bf m}_{i}$ and $\phi_{i}$ but keeps only the $\Omega_{n}=0$ mode. It thus includes classical fluctuations of arbitrary magnitudes but no quantum ($\Omega_{n} \neq 0$) fluctuations. 
The resulting model can be thought of as fermions coupled to the spatially fluctuating random background of the classical field ${\bf m}_{i}$. With these approximations, the effective Hamiltonian 
corresponds to a coupled spin-fermion model, which reads as 
\begin{eqnarray}
H_{eff} & = & \sum_{\langle ij\rangle, \sigma}t_{ij}[c_{i\sigma}^{\dagger}c_{j\sigma}+h.c.] + \sum_{\langle \langle ij\rangle \rangle, \sigma}t_{ij}^{\prime}[c_{i\sigma}^{\dagger}c_{j\sigma}+h.c.] + \sum_{i\sigma}(\frac{U}{2}-\mu)\hat n_{i\sigma} \nonumber \\ && - \frac{U}{2}\sum_{i}{\bf m}_{i}.{\bf \sigma}_{i} + \frac{U}{4}\sum_{i}m_{i}^{2}
\end{eqnarray}
where the last term corresponds to the stiffness cost associated with the now classical field ${\bf m}_{i}$ and ${\bf \sigma}_{i}=\sum_{a,b}c_{ia}^{\dagger}{\bf \sigma}_{ab}c_{ib}={\bf s}_{i}$.

The random background configurations of $\{{\bf m}_{i}\}$ are generated numerically via Monte Carlo simulation and obey the Boltzmann distribution,
\begin{eqnarray}
P\{{\bf m}_{i}\} \propto Tr_{c,c^{\dagger}}e^{-\beta H_{eff}}
\end{eqnarray}

For large and random configurations the trace is computed numerically, wherein we diagonalize $H_{eff}$ for each attempted update of ${\bf m}_{i}$ and converge to the equilibrium configuration using Metropolis algorithm. Evidently, the process is numerically expensive and involves an ${\cal O}(N^{3})$ computational cost per update (where $N=3 \times L\times L$ corresponds to the system size), thus the cost per MC sweep is $N^{4}$. We cut down on the computation by using the traveling cluster algorithm, wherein instead of diagonalizing the entire lattice for each attempted update of ${\bf m}_{i}$ we diagonalize a smaller cluster surrounding the update site. The computation cost now scales as ${\cal O}(NN_{c}^{3})$ (where $N_{c}$ is the size of a smaller cluster surrounding the update site),  which is linear in lattice size $N$. This allows us to access large system sizes, which is essential to capture the inhomogeneity of the underlying magnetic phase.

\end{document}